\documentclass[final,epj]{svjour}
\usepackage{graphics}
\usepackage[T1]{fontenc} 
\usepackage{mathtools,slashed}
\usepackage{comment}
\usepackage[caption=false]{subfig}
\usepackage{amssymb}
\usepackage{subcaption}
\usepackage{graphicx}
\usepackage{graphicx}
\usepackage{dcolumn}
\usepackage{bm}

\makeatletter
\renewcommand\makeheadbox{} 
\makeatother

\begin{document}
\title{The Generalized Uncertainty Principle. New Bounds and Trends}
\author{Ezequiel Valero \inst{1}\inst{2}, Hector Gisbert\inst{1} \and Victor Ilisie\inst{1}
}                     
%
%
\institute{Escuela de Ciencias, Ingeniería y Diseño, Universidad Europea de Valencia, Paseo de la Alameda 7, 46010, Valencia, Spain \and Facultat de Física, Universitat de València, Carrer del Dr. Moliner, 50, 46100 Burjassot, Valencia, Spain}
\date{Received: date / Revised version: date}
%
\abstract{
The Heisenberg uncertainty principle is one of the fundamental pillars of quantum mechanics and 
quantum field theory. It is normally introduced by { postulating the commutation} relations $[\hat{x}^i, \hat{p}^j] = i\hbar \delta^{ij}$. However, as suggested by some quantum gravity models and string theory, this basic principle no longer holds true in the presence of a minimal length, possible the Plank length, and modifications of the commutation have been proposed { i.e., of the form $[\hat{x}^\mu, \hat{p}^\nu ] = -i\hbar(1 +  \beta_0 \, \hat{p}^2/\Lambda^2 )\eta^{\mu\nu}$(plus possible additional terms)}. 
In this work we { will} consider the { previous} modified uncertainty principle in terms of an effective field theory, comment upon some theoretical subtleties that are often overlooked in the literature, and constrain, for the first time, the $\Lambda$ scale with the Compton high-energy experimental data. Our findings suggest that high-energy experiments are potentially sensitive to these corrections and could serve as an effective framework for probing possible violations of the Heisenberg uncertainty principle.
\PACS{
      {PACS-key}{discribing text of that key}   \and
      {PACS-key}{discribing text of that key}
     } 
} 
\maketitle
\section{\label{sec:level1}Introduction}

The Generalized Uncertainty Principle (GUP) was first introduced by \cite{Maggiore:1993kv,Kempf_1995,Scardigli_1999} and extensively
analyzed in later studies \cite{Scardigli_1999,PhysRevD.108.066008,Bang:2006va,Bosso:2023aht,Pedram:2011aa,Galan:2007ns,Das:2012rv,Pedram:2011gw,Mignemi:2011wh,Nozari:2012gd,Tedesco:2011iv,Balasubramanian:2014pba,Tkachuk:2013lta,Bosso:2022vlz,Pramanik:2013zy,Bosso:2018uus,Chung:2018btu,Bosso:2020fos,Bosso:2020jay,Hossenfelder_2006,Das_2016,Maccone_2014,Tawfik_2014,Hossenfelder_2013,Nenmeli_2021,Chen_2013,Basilakos_2010,Gecim_2017,Gecim_2018,KONISHI1990276,Chang_2011,Gomes_2022,Das_2008,ali,Marin}. It normally appears in quantum gravity models \cite{Scardigli_1999,Hossenfelder_2013,Basilakos_2010,Gecim_2018,Das_2008,ali,Marin} and string
theory \cite{KONISHI1990276,Chang_2011,Hossenfelder:2012jw} and is a consequence of the presence of a minimal length, which is usually associated
with the Planck length. The GUP is mathematically introduced by modifying the standard
commutation relations with additional terms, i.e.,
\begin{align}
[\hat{x}^\mu, \hat{p}^\nu] = -i\eta^{\mu\nu}(1+\beta \hat{p}^2 + ...) + \gamma \, \hat{p}^\mu \hat{p}^\nu + ... \, ,
\end{align}
in natural units, where $\beta$ and $\gamma$ can be re expressed as 
\begin{align}
\beta = \frac{\beta_0}{\Lambda^2} \, \qquad \gamma=\frac{\gamma_0}{\Lambda^2} \, ,
\end{align}
with dimensionless constants $\beta_0$ and $\gamma_0$, and $\Lambda$, the energy scale where the GUP becomes relevant, and where $\hat{p}^2$ is given by the scalar product
\begin{align}
    \hat{p}^2 \equiv \hat{p}^\mu \hat{p}_\mu \, . 
\end{align}
As mentioned previously, $\Lambda$ is normally considered on the Planck scale, and
therefore all current high-energy and precision experiments are, at least in principle,
insensitive to the presence of these new terms. However, it is legitimate to consider extensions of the Heisenberg principle regardless of its original motivation from quantum gravity or string theory. Even if it constitutes one
of the building blocks of modern physics, it has not yet been probed with high-energy experiments within an effective field theory framework, which is the main goal of our paper.

In the previously mentioned analyses both relativistic and non-relativistic formulations of GUP have been studied, and some collaborations obtained bounds on the $\beta_0/\Lambda^2$ terms within non-relativistic models, or from astrophysical data. These experimental bounds are nicely summarized in \cite{Bosso:2023aht}. The strongest bound is given by $\beta^{-1/2} = \beta_0^{-1/2} \Lambda > 10$ Tev, within a non-relativistic and anisotropic framework \cite{Gomes_2022}
\begin{align}
    [\hat{x}_i, \hat{p}_j] = i\hbar \left[ f(\hat{\mathbf{p}}) \delta_{ij} + g_i(\hat{\mathbf{p}}) \hat{p}_j \right].
    \label{NRM}
\end{align}
However, these bounds are not directly applicable to our analysis as the previous model is non-relativistic and the deformed commuting relations are different from the ones employed in our analysis. All other bounds are weaker by a few orders of magnitude. Thus, in this manuscript, we derive a precise bound on $\beta$ using high-energy experimental data.

The manuscript is organized as follows. In the first part, we present the theoretical framework, emphasizing the differences with previous approaches. The second part is devoted to the derivation of the leading-order modified QED Lagrangian induced by the GUP, along with the corresponding Feynman rules. We then carry out a phenomenological analysis based on high-energy Compton scattering data \cite{Achard_2005}. Finally, we summarize our findings and conclusions.

\subsection{\label{sec:level2}Preliminary considerations}

Many authors have previously performed calculations within the relativistic GUP framework in quantum field theory using different approaches for the calculations of the physical cross sections and other observables \cite{Hossenfelder:2012jw,Bosso:2020fos,Bosso:2020jay,Hossenfelder:2004up,Quesne:2006is,Kober:2010sj,Kober:2011dn,Husain:2012im,Faizal:2017map}. However, the method employed in this manuscript present some differences with previous works, and all the explicit technical details will be given in this section. 

Let us start with the standard commutation relations that can be expressed in covariant form as
\begin{align}
\label{comm1}
[\hat{x}^\mu, \hat{p}^\nu]&=-i\eta^{\mu\nu} \, , \\
\label{comm2}
[\hat{p}^\mu,\hat{p}^\nu] &= [\hat{x}^\mu,\hat{x}^\nu] = 0 \, , 
\end{align}
where $\eta^{\mu\nu}=\{1, -1, -1, -1\}$ is the Minkowski metric and the operator $\hat{p}^\mu$ can be expressed in terms of the covariant derivative as
\begin{align}
\hat{p}^\mu = i\partial^\mu = i \frac{\partial}{\partial x^\mu} = i (\partial_t,  - \vec{\nabla}) \, .
\end{align}
These relations give rise to the standard Heisenberg uncertainty relations i.e., 
\begin{align}
\Delta x^\mu \Delta p^\nu \geq \frac{1}{2} \delta^{\mu\nu} \, ,
\end{align}
with $\delta^{\mu\nu}$ is the usual Kronecker delta{, and where we have considered natural units $\hbar = c = 1$}.

\subsection{\label{sec:citeref}Minimal GUP}
The minimally modified commutation relations from the expression \eqref{comm1}, that respect \eqref{comm2}, meaning that {they} do not give rise to non-commutative geometries that suffer from further
complications \cite{Quesne:2006is,Todorinov:2018arx}, are given by (see Appendix~\ref{app:deformed_poincare} for further details)
\begin{align}
\label{modgup}
[\hat{x}^\mu, \hat{p}^\nu] = -i\eta^{\mu\nu} (1+\beta \hat{p}^2) -  \, 2 i \beta \hat{p}^\mu \hat{p}^\nu,
\end{align}
where $\beta=\beta_0/\Lambda^2$, with $\beta_0$ a dimensionless constant, and $\Lambda$, the scale at which the GUP becomes relevant, as
mentioned previously. Given the modified commutation relation \eqref{modgup}, the field theory cannot be canonically quantized in the usual way. To address this issue, one introduces an auxiliary operator $\hat{k}^\mu$, which satisfies the standard commutation relations,
\begin{align}
[\hat{x}^\mu, \hat{k}^\nu]=-i\eta^{\mu\nu} \, . 
\end{align}
The newly introduced $\hat{k}^\mu$ operator can be expressed in terms of $\hat{p}^\mu$ as 
\begin{align}
\label{kmu}
\hat{p}^\mu = \hat{k}^\mu (1+\beta \hat{k}^2) \, .
\end{align}
This relation can, equivalently, be expressed in terms of the physical momentum $p^\mu$ as follows,
\begin{align}
p^\mu  p_\mu = p^2 = m^2 = k^2(1+\beta \, k^2)^2 \, .
\end{align}

In this work we shall impose the momentum-conservation relations for the physical momenta, i.e., %
\begin{align}
\sum_i p_i^\mu = \sum_f p_f^\mu \, ,
\end{align}
where $i$ and $f$ are indices that run over the initial and final states. This is obviously
not the case for $k^\mu$. The alternative approach, (momentum conservation for $k^\mu$ instead of $p^\mu$) would give rise to the {\it soccer ball problem} \cite{Hossenfelder:2012jw,Hossenfelder:2014ifa}, that would introduce large contributions to macroscopical objects, which is experimentally discarded.

Let us turn our attention again to \eqref{kmu}. Similar to \cite{Todorinov:2018arx,Kober:2010sj}, in order to obtain the corresponding modified Lagrangian and equations of motion, that can be quantized the usual way, one simply has to perform the substitution
\begin{align}
\label{subst}
i\partial^\mu \to i\partial^\mu (1-\beta \, \partial^\nu \partial_\nu) \, .
\end{align}

In the following section, we seek to clarify the procedure of building an effective Lagrangian and how, once the proper substitutions are made, calculations can proceed as if working in physical momentum space, allowing the final results, such as the transition matrix, to depend only on physical momenta, ensuring momentum conservation in a physically meaningful way.

\subsection{Subtleties in building a modified effective Lagrangian and observables}
As a naive example, consider the charged Klein-Gordon field, including an interaction term, expressed initially as
\begin{align}
\mathcal{L} = (\partial^\mu \phi^\dagger) (\partial_\mu \phi) - m^2 \phi^\dagger \phi + \lambda \phi^\dagger \phi(\partial^\mu \phi^\dagger) (\partial_\mu \phi) \, .
\end{align}
As mentioned previously, in order to quantize this Lagrangian within the framework of the Generalized Uncertainty principle, one has to perform the corresponding substitution \eqref{subst}. We obtain
\begin{align}
\mathcal{L} &= \left[\partial^\mu(1-\beta \, \partial^\nu \partial_\nu) \phi^\dagger\right] \Big[\partial_\mu(1-\beta \, \partial^\nu \partial_\nu) \phi\Big] - m^2 \phi^\dagger \phi  \notag \\
& + \lambda \phi^\dagger \phi \Big[\partial^\mu (1-\beta \, \partial^\nu \partial_\nu) \phi^\dagger\Big] \Big[\partial_\mu (1-\beta \, \partial^\alpha \partial_\alpha) \phi\Big] \, .
\end{align}
We can observe that the propagator has been {\it deformed} when compared to its standard
form, and that new interaction terms have been generated. When calculating the transition matrix for a physical process, i.e.,
\begin{align}
\phi(\vec{p}_1) + \phi^\dagger(\vec{p}_2) \to  \phi(\vec{p}_3) + \phi^\dagger(\vec{p}_4) \, ,
\end{align}
where $\vec{p}_i$ ($i=1,2,3,4$) are the physical momenta, we will obtain the corresponding transition matrix (up to multiplicative constants) 
given by the following expression at tree level
\begin{align}
\label{TransM}
\big<\vec{p}_3,\vec{p}_4 \big| & \mathcal{L}_I(x)\big|\vec{p}_1,  \vec{p}_2 \big> = \notag \\ 
& \big< 0 \big|a(\vec{p}_3) \, b(\vec{p}_4) \, \mathcal{L}_I(x)  \, a^\dagger(\vec{p}_1)  \, b^\dagger(\vec{p}_2) \big|0 \big> \, . 
\end{align}

The subtlety, that is not often mentioned in the literature, consists in noticing that the ladder operators were quantized in terms of momenta $k^\mu$, that do not obey momentum conservation. However, as these operators end up depending on physical momenta, four-momentum properties and conservation will hold as usual. In consequence, the terms in the transition matrix \eqref{TransM} will also depend on the physical momenta, an in general, all observables are defined in a physically meaningful way. The same conclusions can be reached using the path-integral approach, and even more clearly, as it eliminates the need to rely on the commutation relations.

Note that the previous example has been overly-simplified, in the sense that we have
considered an imposed interaction Lagrangian, not originated by a gauge interaction, which
would have been more realistic. This will be done in the following section. However, the only purpose of this naive example is to illustrate the approach that we will be taking in this analysis i.e., once the proper substitutions are made in the Lagrangian, one can calculate observables using the standard procedures, but with a deformed
propagator, with new interaction terms and, completely {\it forgetting} that the quantization was performed in a non-physical basis.

In the following, we shall employ the method presented in this section, and properly
introduce the gauge interactions, in order to obtain the leading terms of the modified
QED Lagrangian up to $\mathcal{O}(\beta)$. Furthermore we shall use these $\mathcal{O}(\beta)$ corrections to perform a phenomenological analysis of the Compton effect using high-energy experimental data \cite{Achard_2005}.

\section{Modified QED Lagrangian}
In order to obtain the modified QED interaction terms and propagators, let us start by considering
the free Dirac and photon field Lagrangian i.e., 
\begin{eqnarray} 
\mathcal{L}_0 = \bar{\psi} (i\gamma^\mu \partial_\mu - m)\psi  - \frac{1}{4} F^{\mu\nu}F_{\mu\nu} \, .
\end{eqnarray}
Introducing the transformation \eqref{subst} we take into account the modifications corresponding
to the GUP. The previous expression then transforms into\footnote{Note that we have found a sign difference of the extra photon field term with respect to \cite{Bosso:2020fos}.}
\begin{align}
\label{L0GUP}
\mathcal{L}^{\text{GUP}}_0 &= 
\bar{\psi} \Big(i\gamma^\mu \partial_\mu(1-\beta \square) - m \Big)\psi  \notag\\
&\quad -\frac{1}{4} F^{\mu\nu}F_{\mu\nu} + \frac{\beta}{2} F^{\mu\nu} \Big(\square F_{\mu\nu} \Big) + \mathcal{O}(\beta^2) \, ,
\end{align}
where we have introduced the usual short-hand notation
\begin{align}
\slashed{\partial} \equiv \gamma^\mu \partial_\mu \, , \qquad \qquad \square \equiv \partial^\alpha \partial_\alpha \, .
\end{align}
The terms from \eqref{L0GUP} give rise to the modified Dirac and photon propagators. Introducing the minimal gauge coupling by means of the usual substitution.. 
\begin{align}
\partial_\mu \to D_\mu = \partial_\mu  + i e Q A_\mu \, ,
\end{align}
for the Dirac Lagrangian from expression \eqref{L0GUP}, we obtain the expression of the GUP modified QED Lagrangian that includes the
extra interaction terms.
\begin{align}
\label{QEDG}
\mathcal{L}^{\text{GUP}}_{\text{QED}} &= 
\bar{\psi} \Big(i\gamma^\mu \partial_\mu (1 - \beta \Box) - m \Big)\psi 
- \frac{1}{4} F^{\mu\nu} F_{\mu\nu} \notag \\
&\quad 
+ \frac{\beta}{2} F^{\mu\nu} (\Box F_{\mu\nu}) 
- e Q A_\mu \bar\psi \gamma^\mu \psi \notag \\
&\quad \quad
+ \tilde{\mathcal{L}}_{I}^\beta + \mathcal{O}(\beta^2) \, .
\end{align}
where $\tilde{\mathcal{L}}_{I}^\beta$ is the $\mathcal{O}(\beta)$ interaction Lagrangian containing the newly generated interaction terms. It is given by
\begin{align}
\tilde{\mathcal{L}}_{I}^\beta &=  -i\beta \, \bar{\psi}\gamma^\mu D_\mu D^\rho D_\rho \psi \, .
\label{Lintnh}
\end{align}
One can easily check that the previous expression is gauge invariant, i.e., it is under the simultaneous transformations
\begin{align}
&\psi \; \to \; \psi' = e^{-ieQ\chi(x)}\psi \, , \notag  \\ 
&A_\mu \; \to \; A'_\mu = A_\mu + \partial_\mu \chi(x) \, ,
\end{align}
with $\chi(x)$ an arbitrary differentiable function. The explicit expression of \eqref{Lintnh}
is given by
\begin{align}
\tilde{\mathcal{L}}_{I}^\beta &=  \beta\, e Q\, \bar{\psi} \gamma^\mu \Big[ 
  A_\mu \Box 
  + 2 (\partial_\mu A^\rho)\, \partial_\rho 
  + 2 A^\rho\, \partial_\rho \partial_\mu \notag \\
&\quad 
  + (\partial_\mu \partial_\rho A^\rho) 
  + (\partial_\rho A^\rho)\, \partial_\mu 
\Big] \psi \notag \\
&\quad \quad 
+ i \beta\, (eQ)^2\, \bar{\psi} \gamma^\mu \Big[
  A^\rho A_\rho\, \partial_\mu 
  + 2 A_\mu A^\rho\, \partial_\rho \notag \\
&\quad \quad \quad 
  + 2 A_\rho (\partial_\mu A^\rho)
  + A_\mu (\partial^\rho A_\rho)
\Big] \psi \notag \\
&\quad \quad \quad \quad 
- \beta\, (eQ)^3\, \bar{\psi} \gamma^\mu A_\mu A^\rho A_\rho\, \psi\, ,
\label{Lib}
\end{align}
similar to the results found in \cite{Bosso:2020fos}. Again, we found a sign difference with respect to \cite{Bosso:2020fos} for the $(eQ)^2$ term, and extra terms, proportional to $(\partial_\rho A^\rho)$ that are absent in the previously mentioned analysis. It is, therefore, necessary to point out that this type of terms do not cancel at the operator level. They are zero for on-shell photons only. 

One can trivially notice that the second line of the previous expression is not Hermitian. However, we can easily obtain the Hermitian version of the $\tilde{\mathcal{L}}_{I}^\beta$ interaction Lagrangian by summing the Hermitian conjugate of the complex part (and multiply the sum by the corresponding 1/2 factor). 

Finally, the Hermitian expression for the $\mathcal{O}(\beta)$ interaction Lagrangian, denoted by ${\mathcal{L}}_{I}^\beta$ is given by
\begin{align}
{\mathcal{L}}_{I}^\beta &= \beta\, e Q\, \bar{\psi} \gamma^\mu \Big[
  A_\mu \Box 
  + 2 (\partial_\mu A^\rho)\, \partial_\rho 
  + 2 A^\rho\, \partial_\rho \partial_\mu \notag \\
&\quad 
  + (\partial_\mu \partial_\rho A^\rho)
  + (\partial_\rho A^\rho)\, \partial_\mu
\Big] \psi \notag \\
&\quad \quad 
+ \frac{i}{2} \beta\, (eQ)^2\, \bar{\psi} \gamma^\mu \Big[
  A^\rho A_\rho\, \partial_\mu 
  - \overset{\leftarrow}{\partial}_\mu A^\rho A_\rho \notag \\
&\quad \quad \quad 
  + 2 A_\mu A^\rho\, \partial_\rho 
  - 2 \overset{\leftarrow}{\partial}_\rho A_\mu A^\rho
\Big] \psi \notag \\
&\quad \quad \quad \quad 
- \beta\, (eQ)^3\, \bar{\psi} \gamma^\mu A_\mu A^\rho A_\rho\, \psi\, .
\end{align}

Alternatively, one can initially part with the Hermitian version of the Dirac Lagrangian.
\begin{align}
    \mathcal{L}_D &= \frac{i}{2} \bar{\psi} \gamma^\mu \overset{\leftrightarrow}{\partial}_\mu \psi - m \bar{\psi} \psi \notag \\
     &= \frac{i}{2} \left( \bar{\psi} \gamma^\mu \partial_\mu \psi - (\partial_\mu \bar{\psi}) \gamma^\mu \psi \right) - m \bar{\psi} \psi \, .
\end{align}
After performing the $\partial_\mu \to \partial_\mu (1-\beta \partial^\rho \partial_\rho)$ shift, and switching the ordinary partial derivative with the covariant derivative, one obtains the Hermitian $\mathcal{O}(\beta)$ interaction Lagrangian.
\begin{align}
    {\mathcal{L}}_{I,h}^\beta &= -\frac{i}{2}\beta \, \bar{\psi}\gamma^\mu D_\mu D^\rho D_\rho \psi + \frac{i}{2}\beta \, 
    \bar{\psi}  \overset{\leftarrow}{D}_\mu \overset{\leftarrow}{D} \, ^\rho \overset{\leftarrow}{D}_\rho \gamma^\mu \psi \, , 
\label{Lib2}
\end{align}
where, as the covariant derivative $\overset{\leftarrow}{D}_\mu$ acts on $\bar\psi$, its expression will be given by
\begin{align}
    \overset{\leftarrow}{D}_\mu = \overset{\leftarrow}{\partial}_\mu  - i e Q A_\mu \, ,
\end{align}
and not by $\overset{\leftarrow}{\partial}_\mu  + i e Q A_\mu$ in order to maintain gauge invariance.\footnote{ As a cross check, if this sign difference with respect to $D_\mu$ is not introduced, one does not obtain the standard QED interaction term. One can alternatively interpret this sign difference the following way: $D_\mu$ couples to $\psi$ with electric charge $Q_\psi = Q$, and 
$\overset{\leftarrow}{D}_\mu$ couples to $\bar{\psi}$ with the opposite electric charge $Q_{\bar\psi} = -Q$.} However, one can check that the expressions \eqref{Lib} and \eqref{Lib2} are equivalent i.e.,
\begin{align}
     {\mathcal{L}}_{I,h}^\beta =  {\mathcal{L}}_{I}^\beta \, ,
\end{align}
up to a total derivative.

\subsection{Reduced interaction Lagrangian and Feynman rules}
\begin{figure*}[t]
\centering
\includegraphics[scale=0.52]{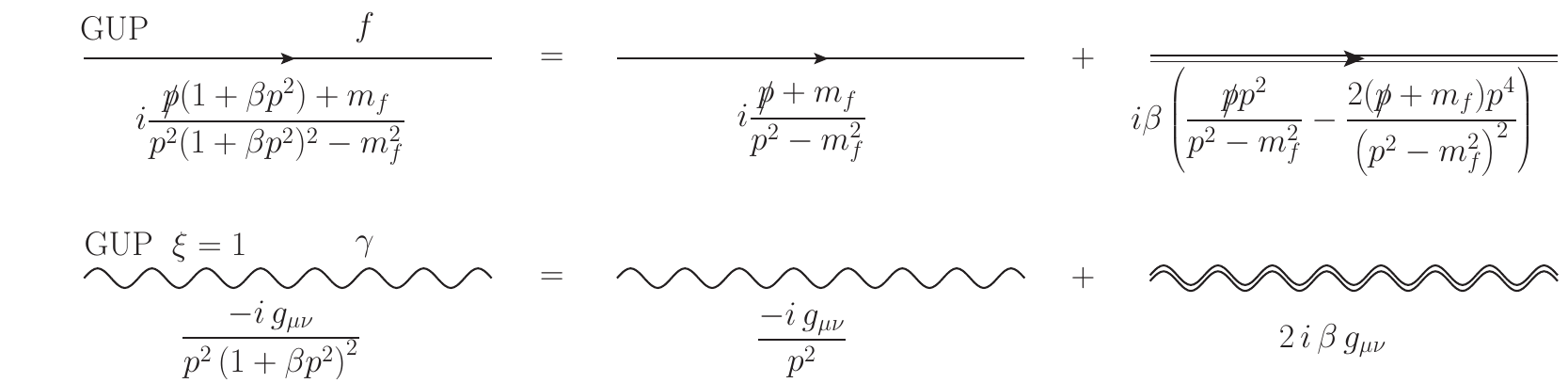}
\caption{Modified Feynman rules for the propagators expressed as a sum of the SM contributions and the $\mathcal{O}(\beta)$ corrections.} 
\label{prop}
\end{figure*} 
One can further reduce the complete interaction Lagrangian $\mathcal{L}_I^\beta$ obtained in \eqref{Lib} by using the equations of motion, i.e., 
\begin{align}
    \frac{i}{2}\bar{\psi} \gamma^\mu  \left( D_\mu  -  \overset{\leftarrow}{D}_\mu \right) \psi = m\bar{\psi}\psi + \mathcal{O} (\beta) \, .
\label{eqmov}
\end{align}
By joining the second and the third line of \eqref{Lib} we get
\begin{align}
{\mathcal{L}}_{I}^\beta &= \beta\, e Q\, \bar{\psi} \gamma^\mu \Big[
    A_\mu \Box
    + 2 (\partial_\mu A^\rho)\, \partial_\rho
    + 2 A^\rho\, \partial_\rho \partial_\mu \notag \\
&\quad
    + (\partial_\mu \partial_\rho A^\rho)
    + (\partial_\rho A^\rho)\, \partial_\mu
\Big] \psi \notag \\
&\quad \quad 
+ \frac{i}{2} \beta\, (eQ)^2\, \bar{\psi} \gamma^\mu \Big[
    A^\rho A_\rho\, D_\mu
    - \overset{\leftarrow}{D}_\mu A^\rho A_\rho \notag \\
&\quad \quad \quad 
    + 2 A_\mu A^\rho\, \partial_\rho
    - 2 \overset{\leftarrow}{\partial}_\rho A_\mu A^\rho
\Big] \psi\, .
\end{align}

and by applying \eqref{eqmov} we finally obtain
\begin{align}
{\mathcal{L}}_{I}^\beta &= \beta\, e Q\, \bar{\psi} \gamma^\mu \Big[
    A_\mu \Box
    + 2 (\partial_\mu A^\rho)\, \partial_\rho
    + 2 A^\rho\, \partial_\rho \partial_\mu \notag \\
&\quad
    + (\partial_\mu \partial_\rho A^\rho)
    + (\partial_\rho A^\rho)\, \partial_\mu
\Big] \psi \notag \\
&\quad \quad 
+ \frac{i}{2} \beta\, (eQ)^2\, \bar{\psi} \gamma^\mu \Big[
    2 A_\mu A^\rho\, \partial_\rho
    - 2 \overset{\leftarrow}{\partial}_\rho A_\mu A^\rho
\Big] \psi\,\notag \\
&\quad \quad  \quad 
+ m\, \beta\, (eQ)^2\, A^\rho A_\rho\, \bar{\psi} \psi  .
\label{LIBB}
\end{align}

which is the interaction Lagrangian that we will be using for our calculations. 

\begin{figure*}[t]
\centering
\includegraphics[scale=0.48]{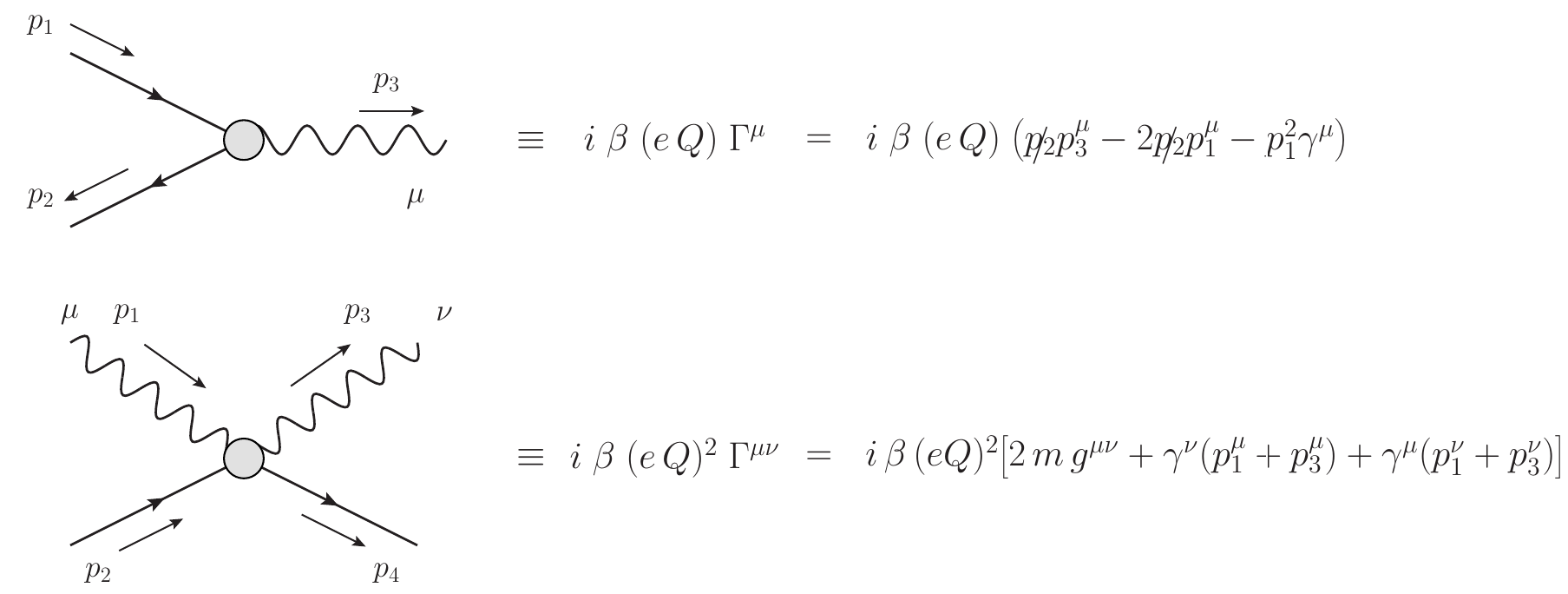}
\caption{Feynman rules of the new QED vertices at $\mathcal{O}(\beta)$ where the momentum flow is given by the corresponding arrows. Momentum conservation reads $p_1=p_2+p_3$ for the 3-point vertex and, $p_1+p_2=p_3+p_4$ for the 4-point vertex.} 
\label{vert}
\end{figure*}

The derivation of the Feynman propagators for the SM fields (scalar, fermionic and gauge
fields) can be found in Appendix~\ref{App1}. Ignoring the $i\epsilon$ term, the Feynman rules for the
modified propagators are shown in Figure~\ref{prop}, where we have split the full propagators
(labeled as GUP) into the {\it ordinary} Standard Model contributions and, the new $\mathcal{O}(\beta)$ contributions. The new vertices at $\mathcal{O}(\beta)$ are given in Figure~\ref{vert}.\footnote{For on-shell photons, the term proportional to $p_3^\mu$, from the $\Gamma^\mu$ vertex, vanishes due to the orthogonality condition $\epsilon^r_\mu(p_3) \, p_3^\mu = 0$, where $\epsilon_\mu^3(p_3)$ are the photon polarization vectors. Similarly, the terms proportional to $p_1^\mu$ and to $p_3^\nu$ also vanish for the $\Gamma^{\mu\nu}$ vertex.}

\section{Compton scattering phenomenology}

\begin{figure*}[t]
\centering
\includegraphics[scale=0.55]{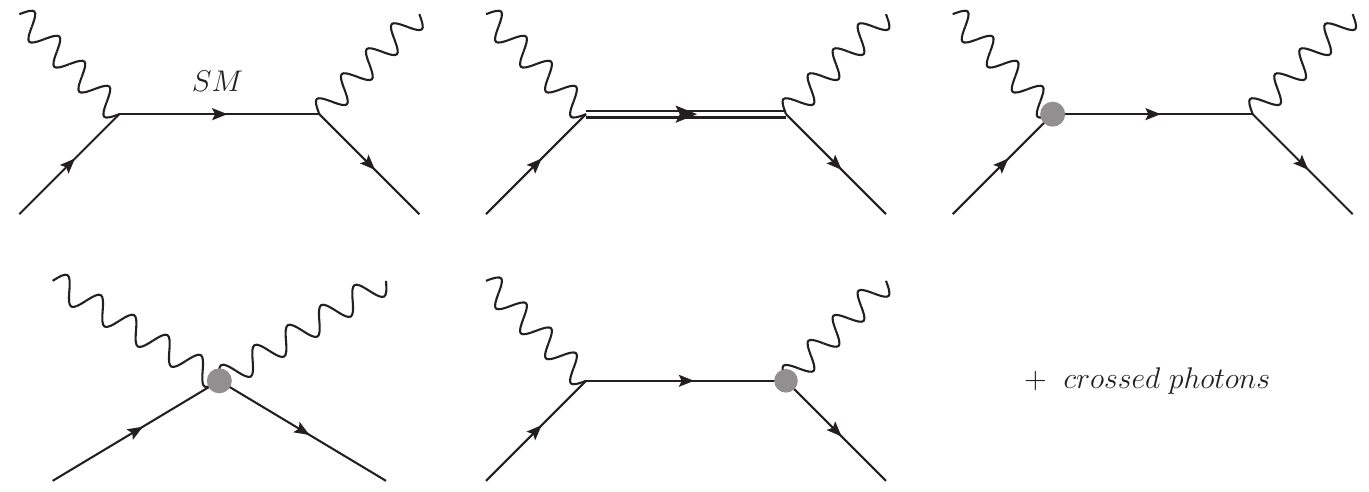}
\caption{Feynman diagrams that contribute to Compton scattering process at  $\mathcal{O}(\beta)$.} 
\label{COMP}
\end{figure*}

The full set of diagrams that contribute to the Compton Scattering cross section at $\mathcal{O}(\beta)$ is shown in Figure~\ref{COMP}. 
In the massless electron limit, the expression of the Compton cross section simply reads
\begin{align}
     \frac{d \sigma_{e\gamma \to e \gamma} }{d \Omega} &= \frac{e^4 \, Q^4}{64 \, \pi^2 s (1+\cos\theta )} \notag \\ 
      & \qquad \times \, \Big( 2 \cos\theta + \cos^2\theta + 5 + 4 \beta s (\cos^2\theta-1) \Big)  \notag \label{Cross}
     \\[1.5ex] &=\frac{d \sigma_{e\gamma \to e \gamma}^{\text{SM}} }{d \Omega} + \alpha^2 \beta \, Q^4 (\cos\theta -1) \,,
\end{align}
where $Q_e = -1$ is the charge of the electron, $e$ the QED coupling constant, and $\alpha$, the fine structure constant. The full expression of the differential cross section including mass terms, is given in Appendix~\ref{fullCross}.

The previous expression is valid in the high-energy limit, $s \gg m^2$, and is therefore suitable for the phenomenological analysis presented below, where we make use of experimental data from LEP \cite{Achard_2005}. It is worth noting that, unlike the (SM) contribution, which scales as $1/s$, the GUP contribution remains constant. This feature makes high-energy experiments particularly promising for probing such new effects, as the SM background becomes increasingly suppressed while the GUP signal persists.

For the phenomenological analysis we define the $\chi^2$ estimator as usual,
\begin{align}
    \chi^2 (\beta)= \sum_{s_j} \frac{\Big(\sigma_{e\gamma\to e\gamma}(s_j,\beta)-\sigma_{e\gamma\to e\gamma}^{\text{exp}}(s_j)\Big)^2}{\mu^2_{\text{exp}}(s_j)} \, ,
\end{align}
where $s_j$ are the different values of $s$ employed in the experimental analysis, $\sigma_{e\gamma\to e\gamma}(s_j,\beta)$ is the theoretical expression for the cross section \eqref{Cross} integrated in between $\cos\theta=-0.8$ and $\cos\theta=0.8$, $\sigma_{e\gamma\to e\gamma}^{\text{exp}}(s_j)$ is the experimental cross section integrated in between the same values of $\cos\theta$ and finally, $\mu^2_{\text{exp}}(s_j)$ is the experimental error (standard deviation). 

\begin{figure}[t]
\centering
\includegraphics[scale=0.70]{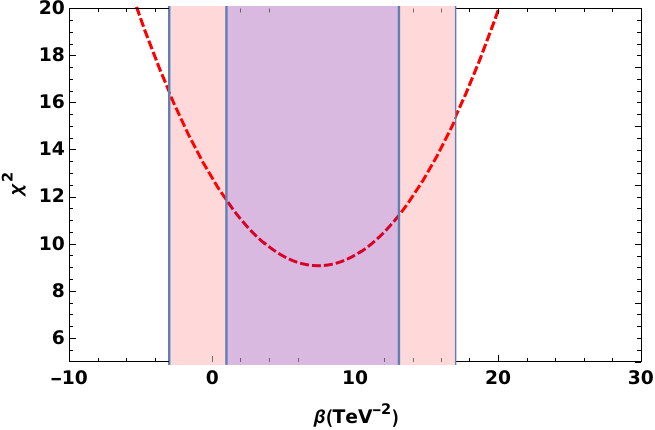}
\caption{$\chi^2$ as function of $\beta$ and the corresponding $1\sigma$ and $2 \sigma$ bandwidths.} 
\label{chi2}
\end{figure}

The results are shown in Figure~\ref{chi2} where we plot the $\chi^2$ as a function of $\beta$ (TeV$^{-2}$). Depending on the sign of the of $\beta_0$, at the $2 \sigma$ level we find the following lower bounds
\begin{align}
   |\beta|^{-1/2} = |\beta_0|^{-1/2} \Lambda > 0.68 \, \text{TeV} \quad \text{with} \,  \beta<0,\\
    |\beta|^{-1/2} = |\beta_0|^{-1/2} \Lambda> 0.25 \, \text{TeV} \quad \text{with}  \, \beta > 0,
    \label{lowbo}
\end{align}
It is worth noting that the energy bounds on $|\beta_0|^{-1/2} \Lambda$ lie below the TeV scale, which makes it particularly appealing to further explore LHC phenomenology within the GUP framework, including potential effects on Higgs-related observables.

Once the $\chi^2$ test has been performed, it is insightful to examine how our model deviates from the SM as a function of the center-of-mass energy ($s$). To carry out this analysis, we integrate over $\cos \theta$ within the angular limits previously defined, and we use the following expression:
\begin{align} 
\Delta \sigma = \frac{\sigma^{\text{SM}} - \sigma^{\text{GUP}}} {\sigma^{\text{SM}}} 
\end{align}
\begin{figure}[t]
\centering
\includegraphics[scale=0.6]{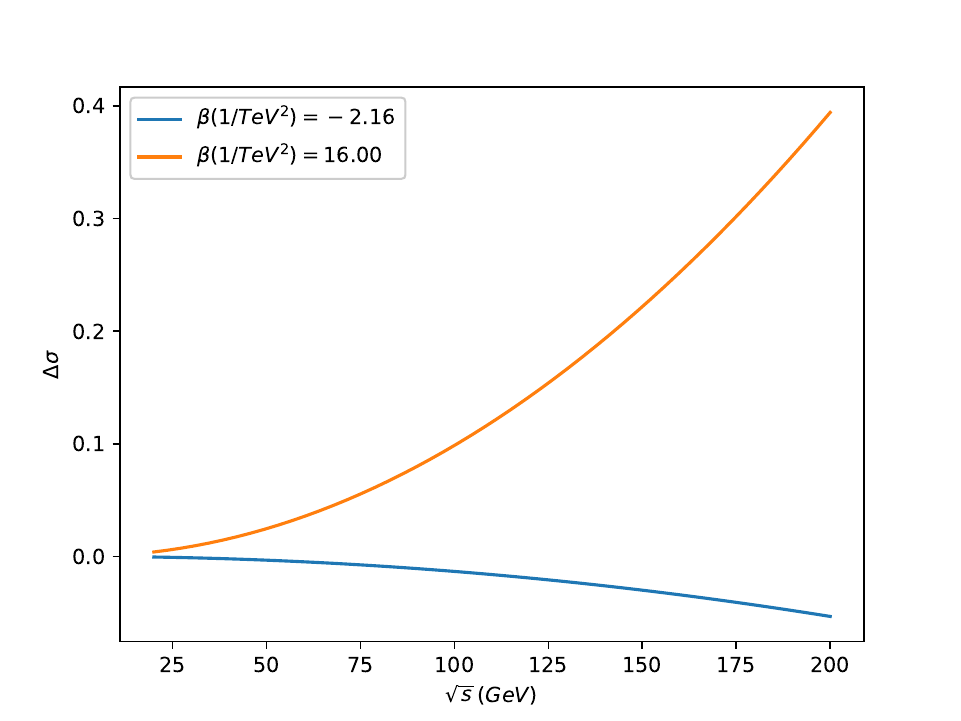}
\caption{$\Delta \sigma$ ($\text{GeV}^{-1}$) as function of $\sqrt{s}$ for two values of $\beta$ that correspond to the lower bounds \eqref{lowbo}.} 
\label{DSigma}
\end{figure} 
The results for the upper and lower bounds of the parameter $\beta$ are shown in Figure~\ref{DSigma}.
\begin{figure*}[t!]
\centering
\subfloat[$\sqrt{s} = 100$ GeV]{
  \includegraphics[width=0.48\textwidth]{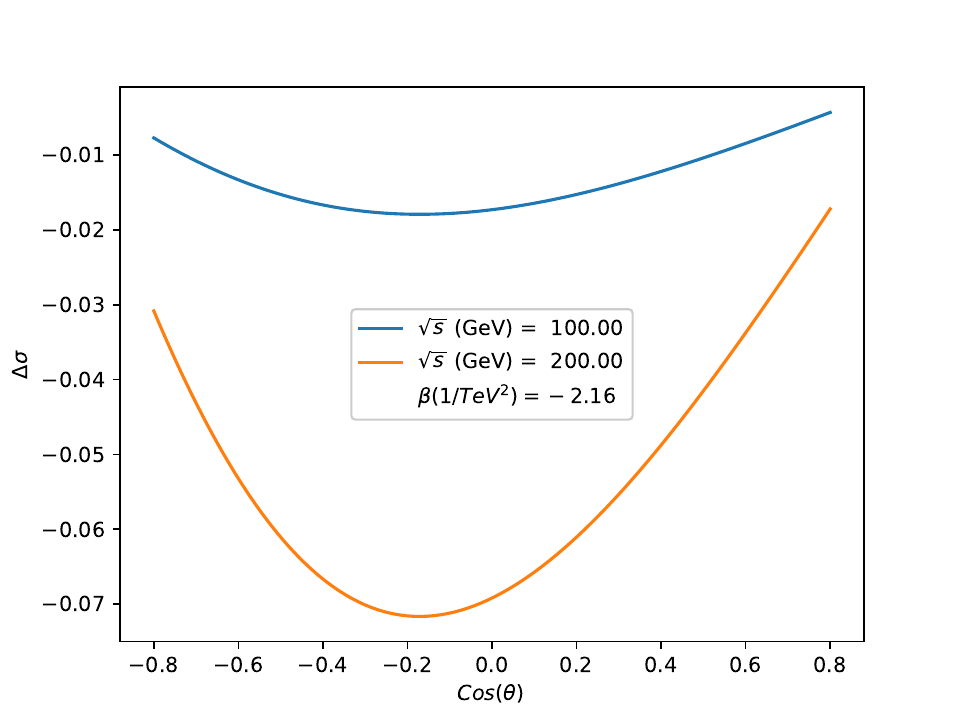}
  \label{fig:plot2}
}
\hfill
\subfloat[$\sqrt{s} = 200$ GeV]{
  \includegraphics[width=0.48\textwidth]{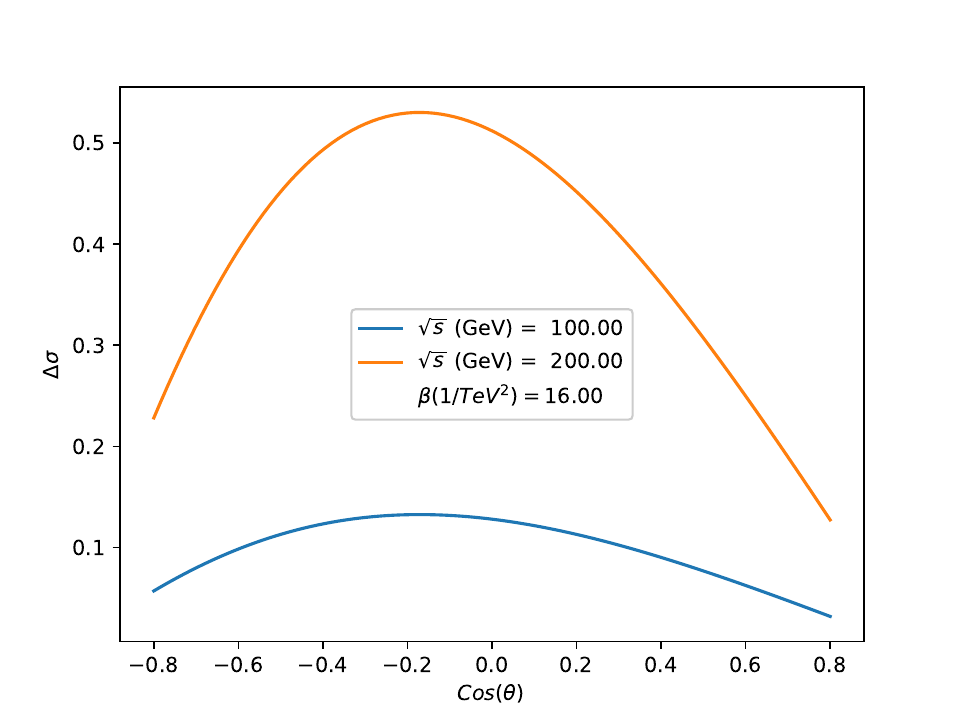}
  \label{fig:plot3}
}
\caption{
Deviation between the Standard Model and the GUP prediction $\Delta\sigma$ as a function of $\cos\theta$ for two energy values, using the lower bounds on $\beta$ from Eq.~\eqref{lowbo}.}
\label{fig:comparacion}
\end{figure*}

From the initial analysis, we observe that the positive bound leads to larger deviations from the SM and becomes relevant at lower energies than the negative bound. Specifically, for the observable associated with the Compton scattering process, the contribution from the Generalized Uncertainty Principle at leading order $\mathcal{O}(\beta)$ reaches approximately $40\%$ near an energy scale of 200 GeV.

It is also instructive to analyze the angular dependence of the GUP contribution at high energies. As illustrated in Figure~\ref{fig:comparacion}, the largest discrepancies between the SM and GUP predictions occur at incidence angles close to $\pi/2$ (low values of $\cos \theta$). For the negative $2\sigma$ bound, the GUP contribution differs from the SM prediction by around $10\%$, whereas for the positive $2\sigma$ bound, the deviation increases to nearly $50\%$. These results suggest that, at an energy scale of 200 GeV, the effects introduced by the modified commutation relations may be experimentally observable in the angular distribution of the cross section.


\section{Conclusions}

In this work, we have presented a consistent formulation of Quantum Electrodynamics within the framework of the Generalized Uncertainty Principle, incorporating the effects of a modified set of commutation relations, encoded in a free parameter $\beta$. Starting from first principles, we have shown that maintaining both gauge and Lorentz invariance requires the inclusion of an additional $k$-space where the commutation relations hold as usual. This leads to a modified, consistent, effective field theory with new vertices and propagators contributing at $\mathcal{O}(\beta)$.

Using this effective GUP-QED Lagrangian, we performed a detailed phenomenological analysis of the Compton scattering process. Our results show that the GUP contributions lead to measurable deviations from the Standard Model predictions, with discrepancies of up to $40\%$ at energies around 200 GeV for the positive $2\sigma$ bound. Angular analyses further reveal that such deviations can reach $50\%$ at low incidence angles, highlighting the potential for experimental detection.

Importantly, the derived bounds on $\beta$ lie below the TeV scale, making future studies at the LHC, especially those involving Higgs-related observables, particularly promising in the search for GUP effects that could ultimately be related to effective, or even fundamental, quantum gravity theories. Our results demonstrate that effective field theory techniques can be successfully employed to bridge quantum gravity scenarios with high-energy phenomenology, offering a testable path for exploring minimal length effects in current and upcoming collider experiments.

The results obtained in this study motivate further investigation of GUP effects in high-energy processes at current and future colliders. In particular, while our analysis has focused on energy scales up to a few hundred GeV, the LHC and proposed future facilities such as the Future Circular Collider (FCC) offer access to significantly higher energies and improved experimental precision. However, the FCC-ee phase \cite{Suarez:2022pcn,Blondel:2019jmp} is expected to operate at center-of-mass energies ranging from 90 to 365 GeV, which aligns remarkably well with the energy range explored in Figures~\ref{DSigma} and~\ref{fig:comparacion}, making our results directly relevant for future precision tests at the FCC.

The FCC, with a projected center-of-mass energy of up to 100 TeV and expected integrated luminosities surpassing those of the HL-LHC, is anticipated to achieve percent-level precision in key electroweak and QED observables. This level of sensitivity opens the possibility of probing GUP-induced deviations well beyond current limits. Higgs physics, diboson production, and precision QED processes at the FCC-ee phase, in particular, could serve as clean environments to test the presence of GUP-induced effects, especially in angular distributions and differential cross sections.

Future work will focus on extending the effective GUP-QED framework to other processes of phenomenological interest such as Higgs decays, electron and muon $(g-2)$ or loop-level corrections. Furthermore, incorporating renormalization group effects and developing a systematic approach to higher-order $\beta$ corrections could refine the predictions and improve the robustness of the bounds.

\appendix

\section{Deformed phase space, Poincar\'e generators and scattering framework}
\label{app:deformed_poincare}

In this study we assume a commutative space-time and commuting physical momenta,\footnote{We will eliminate the {\it hat} notation for quantum operators for simplicity.}
\begin{align}
[x^\mu,x^\nu] &= 0, \label{A1}\\
[p^\mu,p^\nu] &= 0. \label{A2}
\end{align}
The deformation is introduced through the covariant generalized commutation relation
\begin{align}
[x^\mu,p^\nu]
&= -i \eta^{\mu\nu}(1+\beta p^2)-2i\beta p^\mu p^\nu,
\label{A3}
\end{align}
where
\begin{align}
p^2 \equiv \eta_{\mu\nu}p^\mu p^\nu,
\end{align}
and $\beta =\beta_0 \, \Lambda^{-2}$ is the deformation parameter.  
The standard Heisenberg algebra is recovered in the limit $\beta \to 0$. Space-time translations are generated by the physical momentum operator,
\begin{align}
T(a) = \exp(-i a_\mu p^\mu).
\label{A4}
\end{align}
Using the Baker--Campbell--Hausdorff expansion, the action of a translation on an operator
$A$ is
\begin{align}
T(a)\,A\,T^\dagger(a)
&= A + i[-a_\rho p^\rho , A] \notag \\ & \qquad
+\frac{i^2}{2!}[-a_\rho p^\rho,[-a_\sigma p^\sigma,A]]+\cdots .
\end{align}
Applying this to the spacetime coordinates and using the deformed algebra
\eqref{A3}, one finds
\begin{align}
i[-a_\rho p^\rho , x^\mu]
&= a^\mu(1+\beta p^2)+2\beta (a_\rho p^\rho)\,p^\mu .
\end{align}
Since the physical momenta commute, $[p^\mu,p^\nu]=0$, it follows that
\begin{align}
[-a_\rho p^\rho,\,(1+\beta p^2)] = 0,
\qquad
[-a_\rho p^\rho,\,(a_\rho p^\rho)p^\mu] = 0,
\end{align}
so that all higher nested commutators vanish and the translation action closes
at first order,
\begin{align}
T(a)\,x^\mu\,T^\dagger(a)
&= x^\mu + a^\mu(1+\beta p^2)+2\beta (a_\rho p^\rho)\,p^\mu ,
\\
T(a)\,p^\mu\,T^\dagger(a)
&= p^\mu .
\end{align}

On the other hand, for a multiparticle system, the physical momentum acts additively on the
tensor-product Hilbert space. For instance, in a two-particle system,
\begin{align}
p^\mu_{(1)} &\equiv p^\mu_1 \otimes \mathbf{1},
\qquad
p^\mu_{(2)} \equiv \mathbf{1} \otimes p^\mu_2 ,
\end{align}
with the property
\begin{align}
[p^\mu_{(1)},p^\nu_{(2)}]=0 ,
\end{align}
we obtain that the total translation operator factorizes, i.e., 
\begin{align}
T_{\rm tot}(a)
&= \exp\!\big(-i a_\mu (p^\mu_{(1)}+p^\mu_{(2)})\big) \notag  \\
&= \exp(-i a_\mu p^\mu_{(1)})\,\exp(-i a_\mu p^\mu_{(2)}),
\end{align}
and the generator of translations is the total momentum
\begin{align}
p^\mu_{\rm tot} = \sum_{i} p^\mu_{(i)} .
\end{align}
Translation invariance of the $S$-matrix therefore implies standard momentum
conservation in scattering processes,
\begin{align}
\sum_{\rm in} p^\mu = \sum_{\rm out} p^\mu .
\end{align}

Let us now, finally, consider the Lorentz generators of the modified algebra. The standard expression for $M^{\mu\nu}$ given by
\begin{align}
M^{\mu\nu}_0 = x^\mu p^\nu - x^\nu p^\mu
\end{align}
does not generate standard Lorentz transformations on $p^\mu$ when the deformed commutator \eqref{A3} is used.
To restore the standard Lorentz action on physical momentum, we define the modified generators
\begin{align}
\widetilde M^{\mu\nu}
&\equiv \frac{1}{1+\beta p^2}\,
\big(x^\mu p^\nu - x^\nu p^\mu\big) \notag \\ &= M^{\mu\nu}_0 (1-\beta p^2) + \mathcal{O}(\beta^2).
\label{A8}
\end{align}
Using \eqref{A3}, one finds, at $\mathcal{O}(\beta)$
\begin{align}
[\widetilde M^{\mu\nu},p^\rho]
&= i\big(\eta^{\nu\rho}p^\mu-\eta^{\mu\rho}p^\nu\big),
\label{A9}
\end{align}
so that $p^\mu$ transforms as a standard Lorentz four-vector. In particular,
\begin{align}
[\widetilde M^{\mu\nu},p^2] = 0,
\end{align}
therefore, if we impose the mass shell-condition
\begin{align}
p^2 = m^2 ,
\label{shell}
\end{align}
we can observe that it is a Casimir and Lorentz invariant quantity.
The action of $\widetilde M^{\mu\nu}$ on $x^\rho$ is momentum dependent,
\begin{align}
[\widetilde M^{\mu\nu},x^\rho]
&= i(\eta^{\nu\rho}x^\mu-\eta^{\mu\rho}x^\nu)
+ \mathcal{O}(\beta),
\label{A10}
\end{align}
so spacetime coordinates do not transform as a pure Lorentz four-vector beyond the $\beta\to0$ limit.
 This reflects the fact that Lorentz symmetry is realized nonlinearly in
position space, while remaining exact in momentum space. Such behavior should
not come as a surprise in models inspired by the existence of a minimal length,
where the operational meaning of spacetime coordinates is known to be modified
and localization properties become momentum dependent. In this context,
$x^\mu$ should be regarded as an auxiliary phase-space variable, whereas physical
observables and scattering processes are most naturally formulated in momentum
space.

Since $p^\mu$ transforms as a four-vector and $p^2$ is a Lorentz scalar, the right-hand side of \eqref{A3} is manifestly covariant. 
A finite Lorentz transformation is implemented by the unitary operator
\begin{align}
U(\Lambda)
&=\exp\!\left(-\frac{i}{2}\,\omega_{\mu\nu}\,\widetilde M^{\mu\nu}\right),
\label{A11}
\end{align}
where $\omega_{\mu\nu}=-\omega_{\nu\mu}$ are the real (infinitesimal) Lorentz parameters.
To first order, the Lorentz matrix is
\begin{align}
\Lambda^\mu{}_{\nu}
&=\delta^\mu{}_{\nu}+\omega^\mu{}_{\nu}+\mathcal O(\omega^2),
\qquad
\omega^\mu{}_{\nu}\equiv \eta^{\mu\rho}\omega_{\rho\nu},
\label{A12}
\end{align}
and Lorentz invariance implies
\begin{align}
\Lambda^T \eta \Lambda=\eta
\quad \Longleftrightarrow \quad
\omega_{\mu\nu}=-\omega_{\nu\mu}.
\label{A13}
\end{align}

Under the previous transformations, the transformed operators are given by
\begin{align}
x'^\mu &= U(\Lambda)\,x^\mu\,U^\dagger(\Lambda),
\label{A14}\\
p'^\mu &= U(\Lambda)\,p^\mu\,U^\dagger(\Lambda).
\label{A15}
\end{align}
Using the usual expansion,
\begin{align}
U A U^\dagger
&=A+i\left[\frac12\omega_{\rho\sigma}\widetilde M^{\rho\sigma},A\right] \notag \\
& \; \; +\frac{i^2}{2!}\left[\frac12\omega_{\rho\sigma}\widetilde M^{\rho\sigma},
\left[\frac12\omega_{\alpha\beta}\widetilde M^{\alpha\beta},A\right]\right]+\cdots,
\label{A16}
\end{align}
one obtains the infinitesimal variations
\begin{align}
\delta x^\mu
\equiv x'^\mu-x^\mu
&= i\left[\frac12\omega_{\rho\sigma}\widetilde M^{\rho\sigma},x^\mu\right]
+\mathcal O(\omega^2),
\label{A17}\\
\delta p^\mu
\equiv p'^\mu-p^\mu
&= i\left[\frac12\omega_{\rho\sigma}\widetilde M^{\rho\sigma},p^\mu\right]
+\mathcal O(\omega^2).
\label{A18}
\end{align}

By construction of $\widetilde M^{\mu\nu}$ \eqref{A8}, the momentum transforms as a standard Lorentz four-vector,
\begin{align}
[\widetilde M^{\rho\sigma},p^\mu]
&= i\left(\eta^{\sigma\mu}p^\rho-\eta^{\rho\mu}p^\sigma\right),
\label{A19}
\end{align}
and therefore
\begin{align}
\delta p^\mu
&=\omega^\mu{}_{\nu}\,p^\nu +\mathcal O(\omega^2),
\qquad
p'^\mu=\Lambda^\mu{}_{\nu}p^\nu+\mathcal O(\omega^2).
\label{A20}
\end{align}
Finally, the transformed commutator can be written explicitly in terms of the generator as
\begin{align}
[x'^\mu,p'^\nu]
&=\big[U x^\mu U^\dagger,\;U p^\nu U^\dagger\big]
=U\,[x^\mu,p^\nu]\,U^\dagger
\label{A21}\\
&=[x^\mu,p^\nu]
+i\left[\frac12\omega_{\rho\sigma}\widetilde M^{\rho\sigma},[x^\mu,p^\nu]\right]
+\mathcal O(\omega^2). \notag
\end{align}
Using the deformed algebra \eqref{A3}, this implies the form-invariant result
\begin{align}
[x'^\mu,p'^\nu]
&=-i\Big(\eta^{\mu\nu}(1+\beta p'^2)+2\beta p'^\mu p'^\nu\Big),
\label{A23}
\end{align}
with $p'^2=\eta_{\rho\sigma}p'^\rho p'^\sigma=p^2$.

Because translations are generated by the additive physical momentum $p^\mu$, scattering amplitudes satisfy standard momentum conservation,
\begin{align}
(2\pi)^4\delta^{(4)}\!\left(\sum_{\rm in} p^\mu-\sum_{\rm out} p^\mu\right).
\end{align}
External states are placed on the physical mass shell \eqref{shell}, and Lorentz invariance of amplitudes is ensured by building interaction vertices from Lorentz scalars constructed out of $p^\mu$.
All deformation effects can therefore be encoded in higher-dimensional, momentum-dependent operators suppressed by $\Lambda$, while standard momentum-space scattering theory remains applicable within the regime
\begin{align}
|p^2| \ll \Lambda^2.
\end{align}

\section{Calculation of the Modified Feynman Propagators}
\label{App1}

The Feynman propagator $\Delta_F(x)$ for the modified Klein-Gordon field must satisfy
\begin{align}
\Big( \square(1-\beta\square)^2 + m^2 \Big)\Delta_F(x) = - \delta^{(4)}(x) \, . 
\end{align}
The solution can be found straightforwardly i.e., it is simply given by
\begin{align}
\Delta_F(x) = \int \frac{d^4p}{(2\pi)^4} \frac{e^{-ipx}}{p^2(1+\beta p^2)^2-m^2+i\epsilon} \, ,
\end{align}
and so, the Klein-Gordon field propagator will be given, in momentum space, by
\begin{align}
\frac{i}{p^2(1+\beta p^2)^2-m^2} = \frac{i}{p^2-m^2} + \beta \frac{-2 \, i \, p^4}{(p^2 -m^2)^2} + \mathcal{O}(\beta^2) \, .
\end{align}
In the previous equation we have ignored the $i\epsilon$ term, and we shall also do so in the following. The fermionic Feynman propagator $S_F(x)$ must satisfy the modified equation
\begin{align}
S_F(x) = \Big( i \slashed{\partial} (1-\beta\square) + m  \Big) \Delta_F(x) \, ,
\end{align}
and so, we can easily find the expression for the propagator to be
\begin{align}
&\frac{i\left(\slashed{p}(1 + \beta p^2) + m \right)}{p^2(1 + \beta p^2)^2 - m^2}
= \frac{i(\slashed{p} + m)}{p^2 - m^2}  \notag \\ & \qquad
+ i\beta \left[
    \frac{\slashed{p} \, p^2}{p^2 - m^2}
    - \frac{2p^4(\slashed{p} + m)}{(p^2 - m^2)^2}
\right] 
+ \mathcal{O}(\beta^2) \,.
\end{align}
In order to find the Feynman propagator of the photon field, let us take a closer look at
the photon kinetic terms from \eqref{QEDG}. If we chose a an appropriate gauge fixing term, i.e., (one can easily check that the following expression, except the gauge-fixig term, is gauge-invariant)
\begin{align}
\mathcal{L}^{\text{GUP}}_{A,\xi} &= -\frac{1}{4}F^{\mu\nu}F_{\mu\nu} + \frac{\beta}{2} F^{\mu\nu} \Big(\square F_{\mu\nu} \Big) \notag \\ & 
\qquad - \frac{1}{2\xi} \Big(\partial_\mu (1-\beta \square) A^\mu \Big)^2 \, ,
\end{align}
we obtain the following equation of motion for the photon field
\small{
\begin{align}
O^{\mu\nu}_\xi A_\nu = 0 \, .
\end{align}}
with $O^{\mu\nu}_\xi$ defined as
\begin{align}
O^{\mu\nu}_\xi  \equiv  \left( g^{\mu\nu}\square(1-\beta\square)^2 + \left(\frac{1}{\xi} - 1 \right)\partial^\mu\partial^\nu (1-\beta\square)^2 \right)  .
\end{align}
The Feynman propagator must then satisfy
\begin{align}
O^{\mu\nu}_\xi \, D^F_{\nu\alpha}(x) = \delta^\mu_\alpha \delta^{(4)}(x) \, . 
\end{align}
Choosing the Feynman gauge $\xi=1$ we can straightforwardly obtain
\begin{align}
D^F_{\mu\nu} = \int \frac{d^4p}{(2\pi)^4} \frac{-g_{\mu\nu} \, e^{-ipx}}{p^2(1+\beta p^2)^2 + i\epsilon} \, ,
\end{align}
which gives rise to the simple Feynman rule
\begin{align}
\frac{-i \, g_{\mu\nu}}{p^2(1+\beta p^2)^2} = -i \, \frac{g_{\mu\nu}}{p^2} + 2 \, i \, \beta \, g_{\mu\nu} \, .
\end{align}

Let us finally turn our attention to the massive gauge fields $V$ (with $V = W, Z$). The
corresponding GUP-modified equation, in the unitary gauge, will be given by
\begin{align}
O_V^{\mu \nu} V_\nu=0  ,
\end{align}
with $O_V^{\mu \nu}$ 
\begin{align}
O_V^{\mu \nu} \equiv \Big(g^{\mu \nu}\left(\square(1-\beta \square)^2+m^2\right)-\partial^\mu \partial^\nu(1-\beta \square)^2\Big) \, ,
\end{align}
and so, the equation that the propagator of the gauge fields $V$ must satisfy is
\begin{align}
O_V^{\mu \nu} D_{\mu \alpha}^{F, V}(x)=\delta_\alpha^\mu \delta^{(4)}(x) \, ,
\end{align}
and so, the propagator, in the unitary gauge and in momentum space, up to $\mathcal{O}(\beta)$, is simply given by the following expression
\begin{align}
&\frac{1}{p^2-m^2}\left(-g_{\mu \nu}+\frac{p_\mu p_\nu}{m^2}\right) + \notag \\ 
& \qquad \frac{2 \beta \, p^4}{\left(p^2-m^2\right)^2}\left(g_{\mu \nu}-\frac{p_\mu p_\nu}{p^2}\right)+\mathcal{O}\left(\beta^2\right) \, .
\end{align}

\section{Full Expression of the Compton Cross Section}
\label{fullCross}

The unpolarized, spin-averaged, squared amplitude at $\mathcal{O}(\beta)$ for the process is given by the sum 
\begin{align}
    |\overline{\mathcal{M}}|^2 = \frac{1}{4} \left( \, |{\mathcal{M}}_{ \text{SM}}|^2 + 2 \, \text{Re} [\mathcal{M}_{ \text{SM}}^\dagger\,   \mathcal{M}_{ \text{GUP}}]
    \,  \right) , 
\end{align}
with the following expressions for the two contributions
\begin{align}
    |\mathcal{M}_{\text{SM}}|^2 &= \frac{8 e^4 Q^4}{\left(m^2-{s}\right)^2 \left(m^2-{u}\right)^2} \Big(6 m^8-m^4 \left(3 {s}^2+14 {s} {u}+3 \text{u}^2\right) \notag 
    \\ & \;\;
    +  m^2 ({s}+{u}) \left({s}^2+6 {s} {u}+{u}^2\right)-{s} {u} \left({s}^2+{u}^2\right) \Big)
\end{align}
and,
\begin{align}
   2\,\text{Re}\left[\mathcal{M}_{\text{SM}}^\dagger\, \mathcal{M}_{\text{GUP}}\right]
&= \frac{16 \beta e^4 Q^4}{(m^2 - s)^3 (m^2 - u)^3}
\Big(
    -2 s^3 u^3 (s + u) \notag \\ & \quad
    + 22 m^{14}  
    - 2 m^{12} (7s + 9u)
\nonumber \notag \\ & \quad 
    + m^{10} \left( -5s^2 - 46su + 5u^2 \right) \notag \\ & \quad
    + m^8 \left( 7s^3 + 51s^2 u + 35su^2 - u^3 \right)
\nonumber  \notag\\ & \quad
    - 2 m^6 s \left( s^3 + 6s^2 u - 2s u^2 - 2u^3 \right)) \notag \\ & \quad
    + m^4 s u \left( s^3 - 23s^2 u - 31s u^2 - 3u^3 \right)
\nonumber \notag \\ & \quad
    + m^2 s^2 u^2 \left( 3s^2 + 22s u + 5u^2 \right)
\Big) \,.
\end{align}
The expression of the total differential cross-section in the center of mass is given by
\begin{align}
     \frac{d\sigma_{e\gamma \to e \gamma}}{d \Omega} &=  \frac{1}{64 \pi^2 s^2 u} |\overline{\mathcal{M}}|^2 \, ,
\end{align}
where $u$ can be expressed as a function of $\cos\theta$ as usually  
\begin{align}
 u = m^2 -\frac{s}{2} - \frac{(s-m^2)^2}{2s}  \, \cos\theta \, .   
\end{align}

\bibliographystyle{ieeetr}
\bibliography{apssamp.bib}

\begin{thebibliography}{10}

\bibitem{Maggiore:1993kv}
M.~Maggiore, ``{The Algebraic structure of the generalized uncertainty
  principle},'' {\em Phys. Lett. B}, vol.~319, pp.~83--86, 1993.

\bibitem{Kempf_1995}
A.~Kempf, G.~Mangano, and R.~B. Mann, ``Hilbert space representation of the
  minimal length uncertainty relation,'' {\em Physical Review D}, vol.~52,
  p.~1108–1118, July 1995.

\bibitem{Scardigli_1999}
F.~Scardigli, ``Generalized uncertainty principle in quantum gravity from
  micro-black hole gedanken experiment,'' {\em Physics Letters B}, vol.~452,
  p.~39–44, Apr. 1999.

\bibitem{Bang:2006va}
J.~Y. Bang and M.~S. Berger, ``{Quantum Mechanics and the Generalized
  Uncertainty Principle},'' {\em Phys. Rev. D}, vol.~74, p.~125012, 2006.

\bibitem{Bosso:2023aht}
P.~Bosso, G.~G. Luciano, L.~Petruzziello, and F.~Wagner, ``{30 years in: Quo
  vadis generalized uncertainty principle?},'' {\em Class. Quant. Grav.},
  vol.~40, no.~19, p.~195014, 2023.

\bibitem{Pedram:2011aa}
P.~Pedram, ``{New Approach to Nonperturbative Quantum Mechanics with Minimal
  Length Uncertainty},'' {\em Phys. Rev. D}, vol.~85, p.~024016, 2012.

\bibitem{Galan:2007ns}
P.~Galan and G.~A. Mena~Marugan, ``{Canonical Realizations of Doubly Special
  Relativity},'' {\em Int. J. Mod. Phys. D}, vol.~16, pp.~1133--1147, 2007.

\bibitem{Das:2012rv}
S.~Das and S.~Pramanik, ``{Path Integral for non-relativistic Generalized
  Uncertainty Principle corrected Hamiltonian},'' {\em Phys. Rev. D}, vol.~86,
  p.~085004, 2012.

\bibitem{Pedram:2011gw}
P.~Pedram, ``{A Higher Order GUP with Minimal Length Uncertainty and Maximal
  Momentum},'' {\em Phys. Lett. B}, vol.~714, pp.~317--323, 2012.

\bibitem{Mignemi:2011wh}
S.~Mignemi, ``{Classical and quantum mechanics of the nonrelativistic Snyder
  model in curved space},'' {\em Class. Quant. Grav.}, vol.~29, p.~215019,
  2012.

\bibitem{Nozari:2012gd}
K.~Nozari and A.~Etemadi, ``{Minimal length, maximal momentum and Hilbert space
  representation of quantum mechanics},'' {\em Phys. Rev. D}, vol.~85,
  p.~104029, 2012.

\bibitem{Tedesco:2011iv}
L.~Tedesco, ``{Fine Structure Constant, Domain Walls, and Generalized
  Uncertainty Principle in the Universe},'' {\em Int. J. Math. Math. Sci.},
  vol.~2011, p.~543894, 2011.

\bibitem{Balasubramanian:2014pba}
V.~Balasubramanian, S.~Das, and E.~C. Vagenas, ``{Generalized Uncertainty
  Principle and Self-Adjoint Operators},'' {\em Annals Phys.}, vol.~360,
  pp.~1--18, 2015.

\bibitem{Tkachuk:2013lta}
V.~M. Tkachuk, ``{Galilean and Lorentz Transformations in a Space with
  Generalized Uncertainty Principle},'' {\em Found. Phys.}, vol.~46, no.~12,
  pp.~1666--1679, 2016.

\bibitem{Bosso:2022vlz}
P.~Bosso, L.~Petruzziello, and F.~Wagner, ``{The minimal length is physical},''
  {\em Phys. Lett. B}, vol.~834, p.~137415, 2022.

\bibitem{Pramanik:2013zy}
S.~Pramanik and S.~Ghosh, ``{GUP-based and Snyder Non-Commutative Algebras,
  Relativistic Particle models and Deformed Symmetries and Interaction: A
  Unified Approach},'' {\em Int. J. Mod. Phys. A}, vol.~28, no.~27, p.~1350131,
  2013.

\bibitem{Bosso:2018uus}
P.~Bosso, ``{Rigorous Hamiltonian and Lagrangian analysis of classical and
  quantum theories with minimal length},'' {\em Phys. Rev. D}, vol.~97, no.~12,
  p.~126010, 2018.

\bibitem{Chung:2018btu}
W.~S. Chung and H.~Hassanabadi, ``{New generalized uncertainty principle from
  the doubly special relativity},'' {\em Phys. Lett. B}, vol.~785,
  pp.~127--131, 2018.

\bibitem{Bosso:2020fos}
P.~Bosso, S.~Das, and V.~Todorinov, ``{Quantum field theory with the
  generalized uncertainty principle I: Scalar electrodynamics},'' {\em Annals
  Phys.}, vol.~422, p.~168319, 2020.

\bibitem{Bosso:2020jay}
P.~Bosso, S.~Das, and V.~Todorinov, ``{Quantum field theory with the
  generalized uncertainty principle II: Quantum Electrodynamics},'' {\em Annals
  Phys.}, vol.~424, p.~168350, 2021.

\bibitem{PhysRevD.108.066008}
F.~Wagner, G.~Var\~ao, I.~P. Lobo, and V.~B. Bezerra, ``Quantum-spacetime
  effects on nonrelativistic schr\"odinger evolution,'' {\em Phys. Rev. D},
  vol.~108, p.~066008, Sep 2023.

\bibitem{Hossenfelder_2006}
S.~Hossenfelder, ``Interpretation of quantum field theories with a minimal
  length scale,'' {\em Physical Review D}, vol.~73, May 2006.

\bibitem{Das_2016}
S.~Das, M.~P. Robbins, and M.~A. Walton, ``Generalized uncertainty principle
  corrections to the simple harmonic oscillator in phase space,'' {\em Canadian
  Journal of Physics}, vol.~94, p.~139–146, Jan. 2016.

\bibitem{Maccone_2014}
L.~Maccone and A.~K. Pati, ``Stronger uncertainty relations for all
  incompatible observables,'' {\em Physical Review Letters}, vol.~113, Dec.
  2014.

\bibitem{Tawfik_2014}
A.~Tawfik and A.~Diab, ``Generalized uncertainty principle: Approaches and
  applications,'' {\em International Journal of Modern Physics D}, vol.~23,
  p.~1430025, Oct. 2014.

\bibitem{Hossenfelder_2013}
S.~Hossenfelder, ``Minimal length scale scenarios for quantum gravity,'' {\em
  Living Reviews in Relativity}, vol.~16, Jan. 2013.

\bibitem{Nenmeli_2021}
V.~Nenmeli, S.~Shankaranarayanan, V.~Todorinov, and S.~Das, ``Maximal momentum
  gup leads to quadratic gravity,'' {\em Physics Letters B}, vol.~821,
  p.~136621, Oct. 2021.

\bibitem{Chen_2013}
D.~Chen, H.~Wu, and H.~Yang, ``Fermion’s tunnelling with effects of quantum
  gravity,'' {\em Advances in High Energy Physics}, vol.~2013, p.~1–6, 2013.

\bibitem{Basilakos_2010}
S.~Basilakos, S.~Das, and E.~C. Vagenas, ``Quantum gravity corrections and
  entropy at the planck time,'' {\em Journal of Cosmology and Astroparticle
  Physics}, vol.~2010, p.~027–027, Sept. 2010.

\bibitem{Gecim_2017}
G.~Gecim and Y.~Sucu, ``The gup effect on hawking radiation of the 2 + 1
  dimensional black hole,'' {\em Physics Letters B}, vol.~773, p.~391–394,
  Oct. 2017.

\bibitem{Gecim_2018}
G.~Gecim and Y.~Sucu, ``Quantum gravity effect on the tunneling particles from
  2 + 1-dimensional new-type black hole,'' {\em Advances in High Energy
  Physics}, vol.~2018, p.~1–7, 2018.

\bibitem{KONISHI1990276}
K.~Konishi, G.~Paffuti, and P.~Provero, ``Minimum physical length and the
  generalized uncertainty principle in string theory,'' {\em Physics Letters
  B}, vol.~234, no.~3, pp.~276--284, 1990.

\bibitem{Chang_2011}
L.~N. Chang, Z.~Lewis, D.~Minic, and T.~Takeuchi, ``On the minimal length
  uncertainty relation and the foundations of string theory,'' {\em Advances in
  High Energy Physics}, vol.~2011, p.~1–30, 2011.

\bibitem{Gomes_2022}
A.~H. Gomes, ``Constraining gup models using limits on sme coefficients,'' {\em
  Classical and Quantum Gravity}, vol.~39, p.~225017, Oct. 2022.

\bibitem{Das_2008}
S.~Das and E.~C. Vagenas, ``Universality of quantum gravity corrections,'' {\em
  Physical Review Letters}, vol.~101, Nov. 2008.

\bibitem{ali}
A.~F. Ali, M.~M. Khalil, and E.~C. Vagenas, ``Minimal length in quantum gravity
  and gravitational measurements,'' 2015.

\bibitem{Marin}
M.~F. e.~a. Marin, Francesco, ``Gravitational bar detectors set limits to
  planck-scale physics on macroscopic variables,'' {\em Nature Physics},
  vol.~9, p.~225017, Oct. 2013.

\bibitem{Hossenfelder:2012jw}
S.~Hossenfelder, ``{Minimal Length Scale Scenarios for Quantum Gravity},'' {\em
  Living Rev. Rel.}, vol.~16, p.~2, 2013.

\bibitem{Achard_2005}
P.~Achard, ``Compton scattering of quasi-real virtual photons at lep,'' {\em
  Physics Letters B}, vol.~616, p.~145–158, June 2005.

\bibitem{Hossenfelder:2004up}
S.~Hossenfelder, ``{Running coupling with minimal length},'' {\em Phys. Rev.
  D}, vol.~70, p.~105003, 2004.

\bibitem{Quesne:2006is}
C.~Quesne and V.~M. Tkachuk, ``{Lorentz-covariant deformed algebra with minimal
  length},'' {\em Czech. J. Phys.}, vol.~56, pp.~1269--1274, 2006.

\bibitem{Kober:2010sj}
M.~Kober, ``{Gauge Theories under Incorporation of a Generalized Uncertainty
  Principle},'' {\em Phys. Rev. D}, vol.~82, p.~085017, 2010.

\bibitem{Kober:2011dn}
M.~Kober, ``{Electroweak Theory with a Minimal Length},'' {\em Int. J. Mod.
  Phys. A}, vol.~26, pp.~4251--4285, 2011.

\bibitem{Husain:2012im}
V.~Husain, D.~Kothawala, and S.~S. Seahra, ``{Generalized uncertainty
  principles and quantum field theory},'' {\em Phys. Rev. D}, vol.~87, no.~2,
  p.~025014, 2013.

\bibitem{Faizal:2017map}
M.~Faizal and T.~S. Tsun, ``{Topological defects in a deformed gauge theory},''
  {\em Nucl. Phys. B}, vol.~924, pp.~588--602, 2017.

\bibitem{Todorinov:2018arx}
V.~Todorinov, P.~Bosso, and S.~Das, ``{Relativistic Generalized Uncertainty
  Principle},'' {\em Annals Phys.}, vol.~405, pp.~92--100, 2019.

\bibitem{Hossenfelder:2014ifa}
S.~Hossenfelder, ``{The Soccer-Ball Problem},'' {\em SIGMA}, vol.~10, p.~074,
  2014.

\bibitem{Suarez:2022pcn}
R.~G. Suarez, ``{The Future Circular Collider (FCC) at CERN},'' {\em PoS},
  vol.~DISCRETE2020-2021, p.~009, 2022.

\bibitem{Blondel:2019jmp}
A.~Blondel {\em et~al.}, ``{Polarization and Centre-of-mass Energy Calibration
  at FCC-ee},'' 9 2019.

\end{thebibliography}
\end{document}